\documentclass[twocolumn,english,superscriptaddress,nofootinbib]{revtex4-1}
\usepackage[T1]{fontenc}
\usepackage[latin9]{inputenc}
\usepackage{babel}
\usepackage{amsmath}
\usepackage{amssymb}
\usepackage{graphicx}
\usepackage{esint}
\usepackage{subscript}
\usepackage[unicode=true,pdfusetitle,
 bookmarks=true,bookmarksnumbered=false,bookmarksopen=false,
 breaklinks=false,pdfborder={0 0 1},backref=false,colorlinks=false]
 {hyperref}
\usepackage{breakurl}

\makeatletter

\providecommand{\tabularnewline}{\\}

\@ifundefined{textcolor}{}
{%
 \definecolor{BLACK}{gray}{0}
 \definecolor{WHITE}{gray}{1}
 \definecolor{RED}{rgb}{1,0,0}
 \definecolor{GREEN}{rgb}{0,1,0}
 \definecolor{BLUE}{rgb}{0,0,1}
 \definecolor{CYAN}{cmyk}{1,0,0,0}
 \definecolor{MAGENTA}{cmyk}{0,1,0,0}
 \definecolor{YELLOW}{cmyk}{0,0,1,0}
}

\usepackage{babel}

\usepackage{babel}

\makeatother

\begin{document}

\title{Optical creation and temperature stability of the hidden charge density
wave state in 1\textit{T}-TaS$_{2-x}$Se$_{x}$}

\author{L. Stojchevska}

\affiliation{Complex Matter Dept., Jozef Stefan Institute, Jamova 39, Ljubljana,
SI-1000, Ljubljana, Slovenia}

\author{P. Šutar }

\affiliation{Complex Matter Dept., Jozef Stefan Institute, Jamova 39, Ljubljana,
SI-1000, Ljubljana, Slovenia}

\author{E. Goreshnik }

\affiliation{Dept. of Inorganic Chemistry and Technology, Jozef Stefan Institute,
Jamova 39, 1000 Ljubljana, Slovenia}

\author{D. Mihailovic}

\affiliation{Complex Matter Dept., Jozef Stefan Institute, Jamova 39, Ljubljana,
SI-1000, Ljubljana, Slovenia}

\affiliation{Center of Excellence on Nanoscience and Nanotechnology-Nanocenter
(CENN Nanocenter), Jamova 39, 1000 Ljubljana, Slovenia}

\author{T. Mertelj}

\affiliation{Complex Matter Dept., Jozef Stefan Institute, Jamova 39, Ljubljana,
SI-1000, Ljubljana, Slovenia}

\affiliation{Center of Excellence on Nanoscience and Nanotechnology-Nanocenter
(CENN Nanocenter), Jamova 39, 1000 Ljubljana, Slovenia}

\email{tomaz.mertelj@ijs.si}

\selectlanguage{english}%
\begin{abstract}
The femtosecond transinet optical spectroscopy is employed to study
the relaxation dynamics of the equilibrium and hidden metastable charge-density-wave
states in single crystals of 1\textit{T}-TaS$_{2-x}$Se$_{x}$ as
a function of the Se doping $x$. Similarly to pristine 1\textit{T}-TaS$_{2}$,
the transition to a hidden phase is observed at low temperature after
a quench with a single 50~fs laser pulse, in the commensurate Mott
phase up to $x=0.6$. The photo-induced hidden-phase formation is
accompanied by a notable change in the coherent phonon spectra, and
particularly the collective amplitude mode. While the stability of
the hidden phase with increased temperatures is only slightly dependent
of the Se content the hidden-phase creation-treshold fluence strongly
increases with the Se content from 1 to $\sim4$ mJ/cm$^{2}$.
\end{abstract}
\maketitle

\section*{Introduction}

The interplay of different degrees of freedom shapes the manifold
of emergent electronic and structural ordered phases in low-dimensional
systems with competing interactions. A marked example is the class
of non-semiconducting layered transition-metal dichalcogenides.\cite{Wilson}
In this class 1\textit{T}-TaS$_{2}$ is of particular interest due
to the presence of a meta-stable hidden charge-density-wave state
induced by short strongly-nonequilibrium optical\cite{StojchevskaScience}
or electrical\cite{vaskivskyi2015controlling,hollander2015electrically,vaskivskyi2016fast}
excitations.

In equilibrium the pristine 1\textit{T}-TaS$_{2}$ shows a series
of electronic phase transitions upon cooling from high temperature.\cite{thomson1994scanning}
First, at $T_{\mathrm{IC}}=550$~K a transition to an incommensurate
CDW (IC) phase is observed. With further cooling a first order transition
to a nearly commensurate (NC) phase at $T_{\mathrm{NC}}=350$ K results
in formation of star-shaped-polaron clusters. Below $T_{\mathrm{C}}$=183
K the system undergoes a first-order lock-in transition into a commensurate
(C) phase where the polarons order commensurately with the underlying
atomic lattice. Concurrently, after Brillouin zone folding due to
the enlarged unit cell, the resulting narrow half-filled Ta~5$d$
valence band splits due to the electronic correlations forming a $\sim$~300~meV
Mott insulator gap.\cite{Fazekaz,DardelPRB} Upon heating the C phase
an additional trigonal (T) phase is observed between the C and NC
phases in the range $T_{\mathrm{T}}=220\,\mathrm{K<}T<280\,K$. The
C Mott state can be suppressed by pressure\cite{SiposNature} or chemical
doping\cite{DiSalvoSeDoped} leading to the superconducting\cite{LiuAPL,AngPRB}
ground state. 

The nonequilibrium dynamics of this compound have been under intensive
investigation recently\cite{DemsarPRB,DeanPRL,PerfettiPRL,StojchevskaScience,HauptEichberger2016,LaulheHuber2017,ligges2018ultrafast},
leading to the discovery\cite{StojchevskaScience} of the low-temperature
meta-stable hidden (H) phase. The phase forms under strongly nonequilibrium
conditions on a short timescale\cite{ravnikVaskivskyi2018} and at
low $T$ the H phase is practically stable\cite{StojchevskaScience,vaskivskyi2015controlling}.
With increasing $T$ the characteristic relaxation time, $\tau_{\mathrm{H}}$,
shows activated behavior $\mbox{\ensuremath{\tau}}_{\mathrm{H}}^{-1}\propto\exp(-T_{\mathrm{A}}/T)$,
where $T_{\mathrm{A}}$ depends on the in-plane strain\cite{vaskivskyi2015controlling}.
In all-optical transient reflectivity experiments in bulk samples
$\tau_{\mathrm{H}}$ drops to the single scan timescale of $\sim30$
minutes at $T\sim70$ K.\cite{StojchevskaScience,vaskivskyi2015controlling}
From the point of view of possible memory applications it would be
desirable to improve the stability of the H state at higher $T$.
Furthermore, the origin of metastability is still not fully understood,
and it is thus of importance to investigate the influence of different
external control parameters on its properties. 

In the present case, we introduce isovalent Se substitution for S:
$1T$-TaS$_{2-x}$Se$_{x}$, which exerts a chemical strain and introduces
disorder. The introduction of the chemical strain and disorder strongly
alters the electronic ground state of $1T$-TaS$_{2-x}$Se$_{\ensuremath{x}}$.\cite{LiuAPL,AngPRB}
With increasing $x$ the the hysteresis related to the NC-C-T phase
transitions broadens, with the C phase being pushed to lower $T$
on cooling up to $x\sim0.8$, whereafter the C Mott-insulating ground
state is suppressed and the superconducting ground state appears.\cite{LiuAPL,AngPRB}.
The broader hysteresis at low doping suggests that the doping increases
the free-energy barriers between the C, NC and T phases in addition
to the suppression of the the insulating Mott state. Since a high
enough free-energy barrier is crucial for the stability of the metallic
H state we therefore investigated the influence of the Se doping on
its formation threshold and stability. 

Recently, the collective mode spectral shifts in 1\textit{T}-TaS\textsubscript{2}
were investigated both in equilibrium and in the metastable H and
T states over a large range of temperatures.\cite{ravnikVaskivskyi2018}
Shifts in the collective mode frequency, which presumably arise from
changes of electronic structure accompanying the transition to the
H state, were shown to be a useful fingerprint signature of the H
state. Here we use this to investigate the stability of the H state
in 1\textit{T}-TaS\textsubscript{2-x}Se\textsubscript{x} with different
$x$. 

In the ultrafast transient reflectivity response the low frequency-phonon
and collective mode frequencies are easily characterized by analysing
the coherent transient response. \cite{StojchevskaScience,ravnikVaskivskyi2018}
We therefore employ the ultrafast transient reflectivity spectroscopy
to 1\textit{T}-TaS$_{2-x}$Se$_{x}$ ($x=$ 0, 0.15, 0.2, 0.5 and
0.6) single crystals before and after strong laser pulse photoexcitation
to study the influence of the Se doping on the formation and temperature
stability of the H-phase.

\section*{Experimental}

Single crystals of 1\textit{T}-TaS$_{2-x}$Se$_{x}$ were grown by
means of the chemical transport reaction with iodine as the transport
agent. Appropriate amounts of of Ta, S and Se powders were put together
with a small amount of I$_{2}$ into evacuated quartz ampules and
placed into a three-zone temperature gradient furnace. The crystal
growth was achieved by setting the temperature gradient across the
ampules to $1000\,^{\circ}{\rm C}$$-$$800\,^{\circ}{\rm C}$ for
216~hours. Finally, ampules were quenched into water. 

The Se content was determined by means of the standard energy-dispersive
X-ray spectroscopy (EDS). For the optical measurements the characterized
crystals were mounted on a cold finger of a liquid-He flow optical
cryostat equipped with CaF$_{2}$ windows and cleaved to expose fresh
surface. 

The optical pump-probe experiments were performed using a train of
50~fs laser pulses at 800~nm from a Ti:Sapphire laser system at
a 250-kHz repetition rate. In order to ensure minimal heating and
avoid switching into the H-phase the pump and probe fluences were
kept constant during all measurements and estimated to be $\mathcal{F}_{\mathrm{p}}=15$~$\mu$J/cm$^{2}$
and $\mathcal{F}_{\mathrm{pr}}=0.5$~$\mu$J/cm$^{2}$ for the pump
and probe pulses, respectively.

For switching into the H-phase an additional single laser pulse at
800~nm with fluence $\mathcal{F}_{\mathrm{SW}}=1-5$~mJ/cm$^{2}$
was picked form the pulse train by means of an acousto-optical modulator
driven by a programmable waveform generator. 

\begin{figure}
\includegraphics[width=0.6\columnwidth]{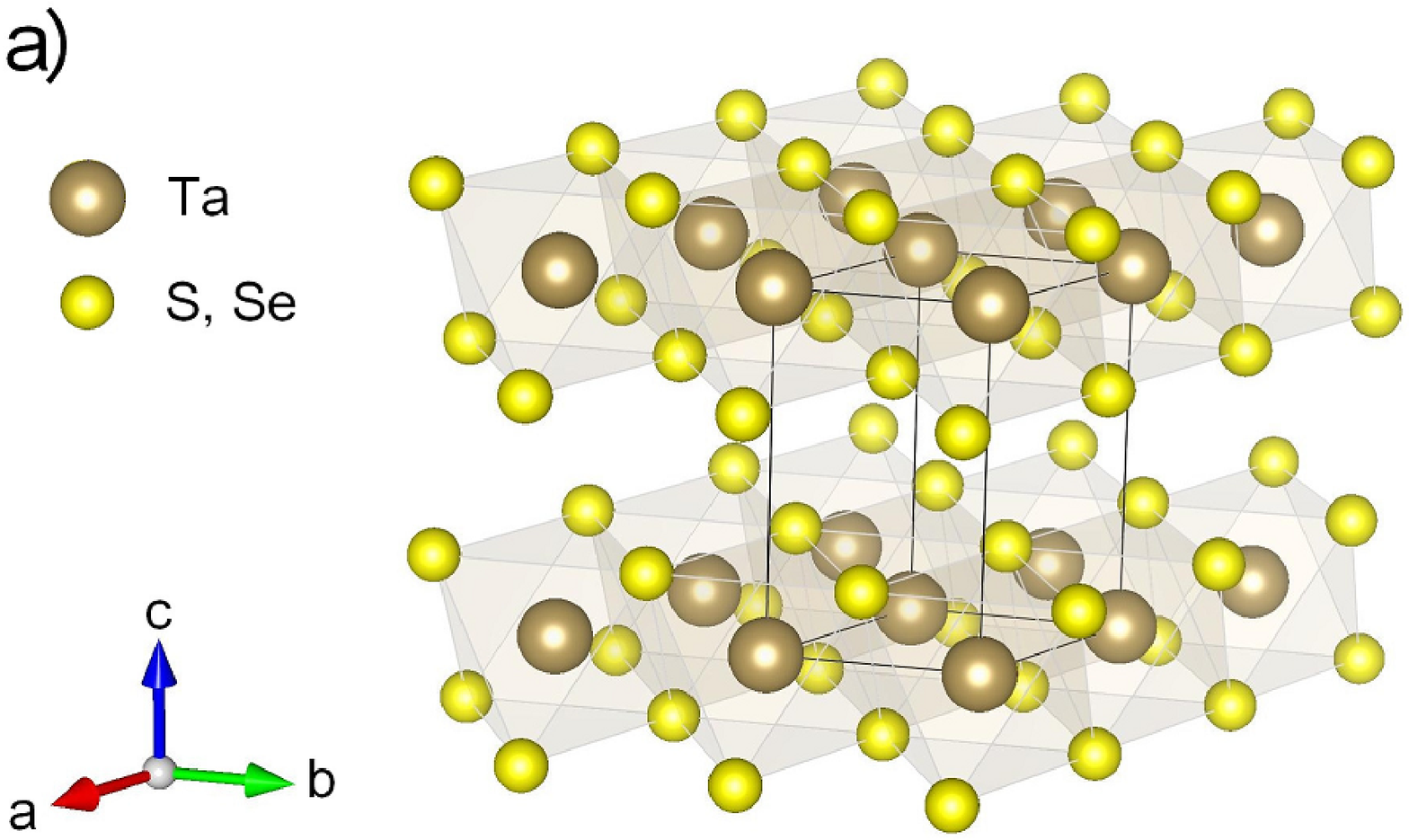} \bigskip{}

\smallskip{}
\includegraphics[width=1\columnwidth]{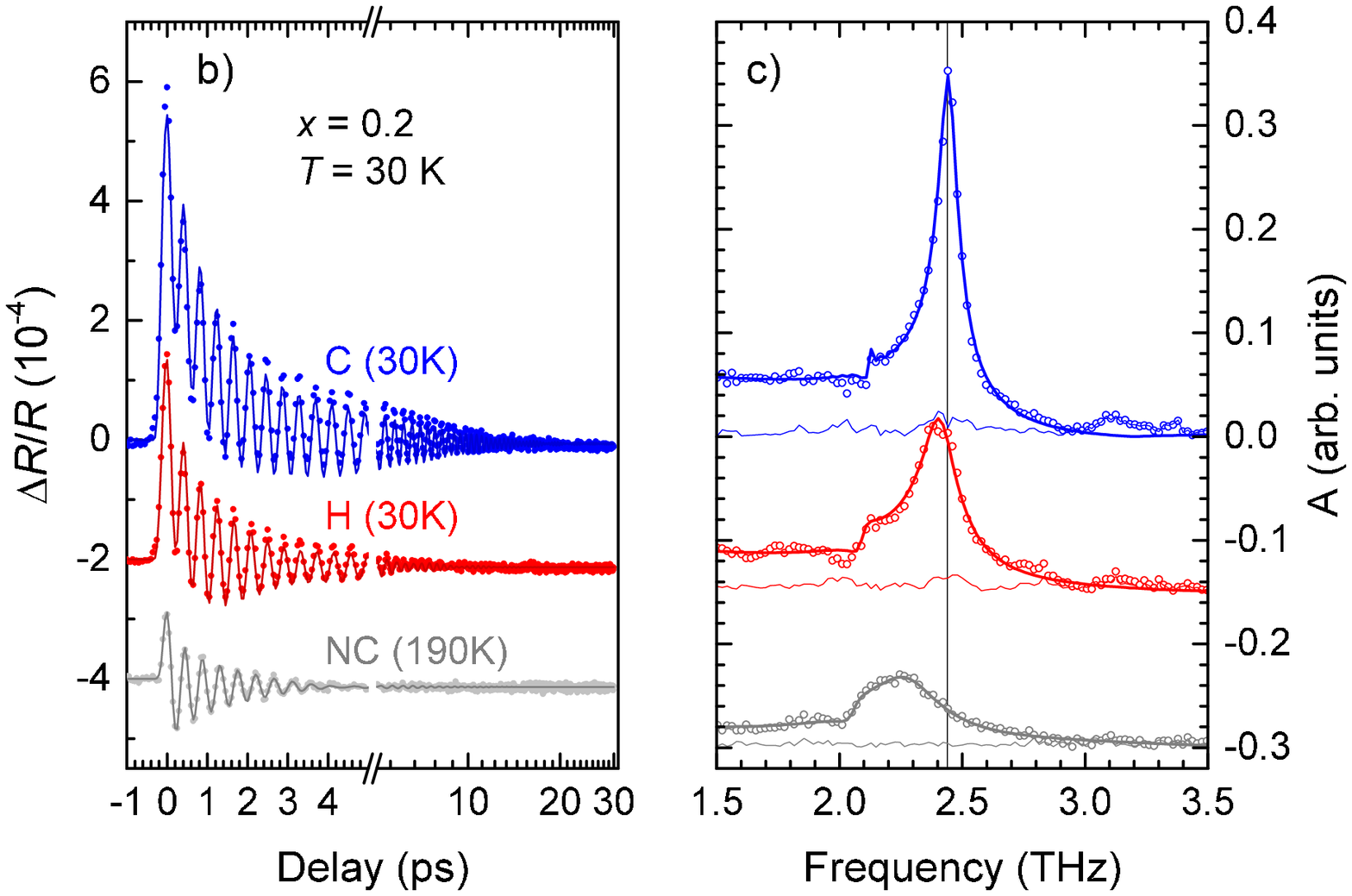}\caption{(a) The crystal structure of 1\textit{T}-TaS$_{2-x}$Se$_{x}$. Transient
reflectivity in different phases (c) with the corresponding FTA spectra
(d). The thick lines are DCE fits. The thin lines in (d) correspond
to the fit residua.}
\label{fig:transients} 
\end{figure}

\begin{figure}[h!]
\begin{centering}
\includegraphics[width=0.6\columnwidth]{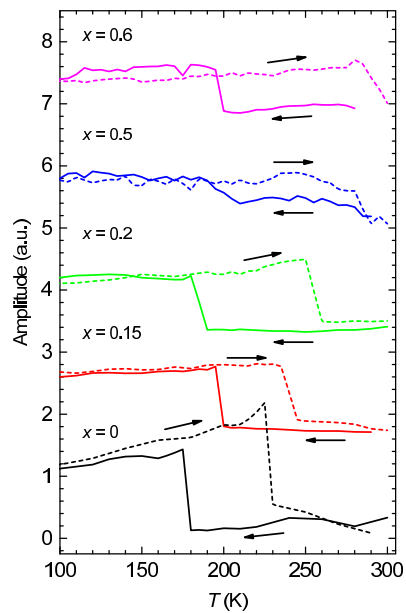} 
\par\end{centering}

\caption{Temperature dependence of the normalized transient reflectivity amplitude
for different $x$. The data for different $x$ are shiftted verticaly
while the arrows indicate cooling and warming scans.}
\label{fig:Hysteresis} 
\end{figure}

\section*{Results}

In Fig. \ref{fig:transients} we plot the charateristic transient
reflectivity transients ($\Delta$$R/R$ ) in the the C, H and NC
phases in the $x=0.2$ sample together with the Fourier transform
amplitude (FTA) spectra. The C and NC phases differ strongly in the
shape and the amplitude of the transients. To determine $T_{\mathrm{C}}$
and $T_{\mathrm{T}}$ we first measured the $T$-dependence of the
transients during a cooling-warming cycle between $300$ K and $\sim70$K
. In Fig. \ref{fig:Hysteresis} we show the normalized amplitude of
the transients. Contrary to Refs. \cite{DiSalvoSeDoped,LiuAPL} we
observe a weak increase of $T_{\mathrm{C}}$ with increasing $x$
while $T_{\mathrm{T}}$ and the broadening of the hysteresis show
a similar trend.

\begin{figure}
\includegraphics[width=0.7\columnwidth]{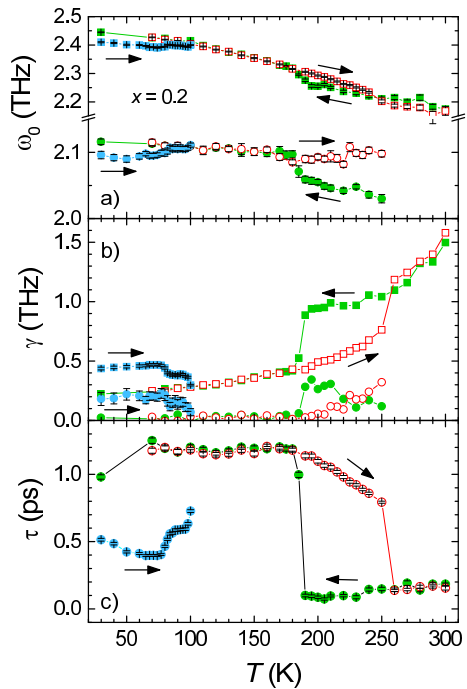}

\caption{Temperature dependence of the modes frequency, $\omega_{0},$ damping,
$\gamma$, and the initial relaxation time, $\tau$, from the DCE
fit for the $x=0.2$ sample. Arrows indicate colling/warming cycles.
The blue symbols correspond to the warming scan in the H-phase after
application of an above-threshold optical switching pulse at $T=30$
K.}
\label{fig:T-dependence}
\end{figure}

We further analyse the $T$-dependent transients using the displacive
coherent excitation (DCE) model\cite{zeiger1992theory,stojchevskaBorovsak2017,ravnikVaskivskyi2018}.
Contrary to $x=0$ \cite{ravnikVaskivskyi2018}, where four modes
were necessary to completely describe the lineshape in the vicinity
of the 2.45-THz amplitude (AM) mode, using only two modes at $\sim2.45$
THz and $\sim2.1$ THz together with the initial exponential relaxation
enabled a fair fit to the data (see Fig. \ref{fig:transients}). The
$T$-dependence of selected fit parameters for $x=0.2$ sample is
shown in Fig. \ref{fig:T-dependence}. The $2.1$-THz mode could be
unambigously fitted only below $T\sim250$ K due to weaker intensities
and strong broadening of the modes in the NC and T phases. Apart from
the frequencies, the mode damping and the initial relaxation decay
time $\tau$, both show notable differences in different CDW phases.

\begin{figure}[h!]
\includegraphics[width=0.7\columnwidth]{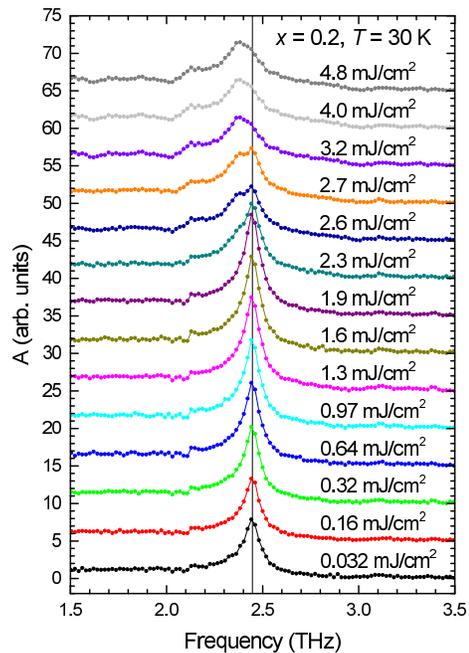} \caption{The switching-pulse fluence dependence of the low-fluence transient
reflectivity FTA spectra at 30~K for $x=0.2$ sample. }
\label{fig:FFT-switching} 
\end{figure}

\begin{figure}[h!]
\includegraphics[width=0.8\columnwidth]{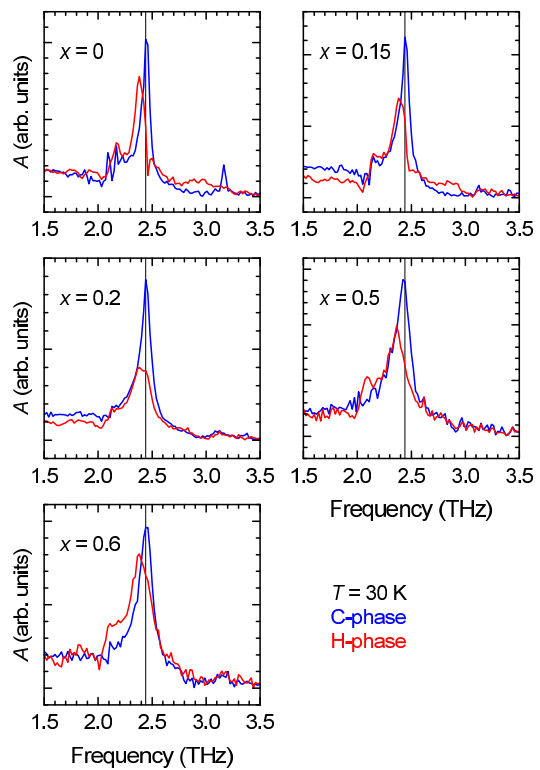} \caption{Reflectivity transients FTA spectra in the C and H-phase at $T=30$
K for different Se doping levels, $x$. }
\label{fig:FFT-all} 
\end{figure}

Next, we checked for the presence of the H-phase and determine the
switching fluence at $T=30$~K by measuring a series of the low-fluence
$\Delta R/R$ transients after an exposure to a single switching pulse
of increasing fluence. An example sequence of the FTA spectra with
increasing $\mathcal{F}_{\mathrm{SW}}$ in the $x=0.2$ sample are
shown in Fig. \ref{fig:FFT-switching}. The H phase is characterized
mainly by the softening of the strongest coherent mode from $\sim2.45$
THz in the C phase to $\sim2.40$ THz (at $T=30$~K). The onset of
the characteristic AM mode spectral weight transfer\cite{StojchevskaScience}
is observed at $\mathcal{F}_{\mathrm{SW}}=2.3$ mJ/cm$^{2}$. The
switching is complete above $\mathcal{F}_{\mathrm{SW}}=2.7$ mJ/cm$^{2}$.
In the intermediate $\mathcal{F}_{\mathrm{SW}}$-interval the spectral
shape indicates incomplete switching with the presence of both, C
and H phases. The behavior is similar for all studied Se contents
with the C- and H- phase spectra and the corresponding threshold fluences,
$\mathcal{F}_{\mathrm{H}}$, shown in Fig. \ref{fig:FFT-all} and
Table \ref{tb:Table}. Here, $\mathcal{F}_{\mathrm{H}}$ is defined
as the minimal fluence where the completed spectral change is observed
and it strongly increases with increasing $x$. 

\begin{table}
\begin{tabular}{ccccc}
$x$  & $T_{\mathrm{C}}$  & $T_{\mathrm{T}}$  & $\mathcal{F}_{\mathrm{H}}$ (@ 30 K)  & $T_{\mathrm{H}}$ \tabularnewline
\hline 
0  & 180~K  & 225~K  & 1~mJ/cm$^{2}$  & 80~K \tabularnewline
0.15  & 200~K  & 240~K  & $1.5\pm0.1$~mJ/cm$^{2}$  & 80~K \tabularnewline
0.2  & 185~K  & 255~K  & $3\pm0.2$~mJ/cm$^{2}$  & 80~K \tabularnewline
0.5  & 200~K  & 280~K  & $4\pm0.4$~mJ/cm$^{2}$  & 95~K\tabularnewline
0.6  & 200~K  & 290~K  & $3.6\pm0.4$~mJ/cm$^{2}$  & 85~K\tabularnewline
\end{tabular}\caption{The transition temperatures to/from the C state upon cooling ($T_{\mathrm{C}}$)
and warming ($T_{\mathrm{T}}$), the threshold fluence, $\mathcal{F}_{H}$,
for switching into the H-phase at $T=30$ K and the highest temperature
$T_{\mathrm{H}}$ up to which the H-phase persists on the timescale
of the pump-probe scan ($\sim30$ min) for different Se dopings. \label{tb:Table}}
\end{table}

Once we had driven the system into the H-phase, we investigated the
$T$ dependence of the reflectivity transients in the H-phase. In
these experiments, the H-phase transition was first triggered by the
above-threshold fluence switching pulse at $T=30$~K. Then the $\Delta$$R/R$
transients were recorded using a weak pump-probe sequence at an increasing
$T$ sequence until the C-phase transient response was observed. 

\begin{figure}
\includegraphics[clip,width=0.8\columnwidth]{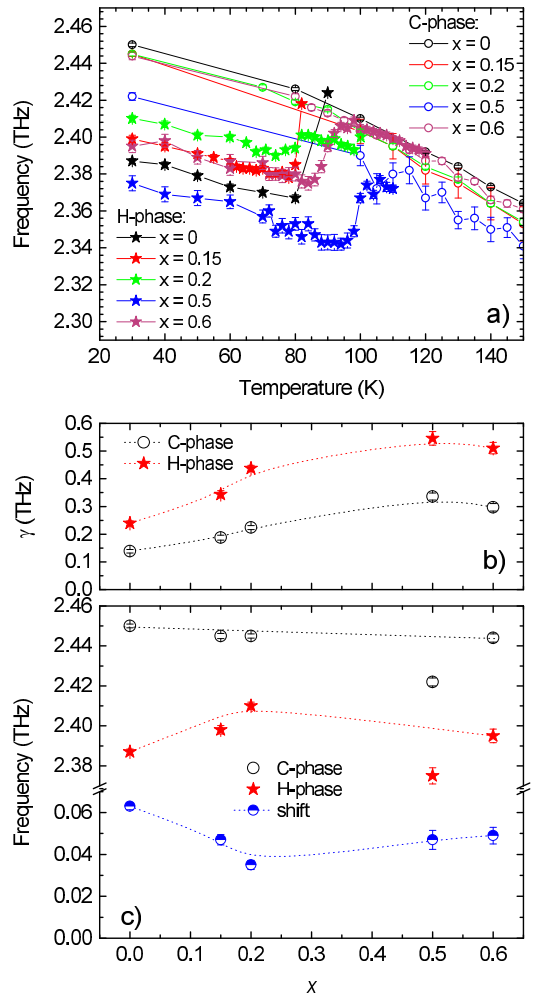}

\caption{(a) Temperature dependence of the 2.40-THz mode frequency on warming
after switching into the H-phase at 30K. The empty symbols refer to
the frequency in the C phase. (b) Se doping dependence of the 2.4(5)-THz
mode damping and (c) the frequency at $T=30$ K in the C and H phases.
The dotted lines are guide lines. The errorbars correspond to the
errors obtained from the DCE-fit.}
\label{fig:Temp-H-State} 
\end{figure}

The $T$-dependence of the selected DCE-fit parameters in the H-phase
for the $x=0.2$ sample is also shown in Fig. \ref{fig:T-dependence}.
While the frequencies of both modes soften in the H-phase, the modes
decay faster (damping, $\gamma$, increases) and the initial exponential
relaxation time, $\tau$, decreases. With increasing $T$ the strongest
H-phase mode shows a similar softening with increasing $T$ as in
the C-phase (see also Fig. \ref{fig:Temp-H-State}) until, at $T_{\mathrm{H}}$,
a recovery of the C-phase transient response is observed.\footnote{Since the stability time, $\tau_{\mathrm{H}}$, of the H-phase strongly
depends on $T$ \cite{vaskivskyi2015controlling} $T_{\mathrm{H}}$
is set by the timescale of experiment\cite{ravnikVaskivskyi2018},
that was $\sim30$ minutes in the present case. On shorter timescale
the H-phase can be readily observed at much higher $T$.\cite{ravnikVaskivskyi2018}} 

$T_{\mathrm{H}}\sim80$ K does not show a strong variation with $x$\footnote{In the $x=0.2$ sample we observe a two step recovery of the C-phase
with the dominant recovery of the spectral shape at $\sim80$ K with
the H-phase signature up to $\sim100$ K. } except for $x=0.5$ with $T_{\mathrm{H}}\sim95$ K (Table \ref{tb:Table}).
In the $x=0.5$ sample we observe also a slightly softer AM mode in
the C-phase together with the softer corresponding mode in the H-phase
while for other dopings the frequency shift is observed only in the
H-phase.

\section*{Discussion}

Contrary to earlier results\cite{DiSalvoSeDoped,LiuAPL} our samples
do not show a significant decrease of $T_{\mathrm{C}}$ with increasing
$x.$ Since the NC$\rightarrow$C transition is of first order the
pinning of the CDW in the NC phase by the Se disorder can play a significant
role in supercooling the NC phase. Since the disorder can be sample
dependent the difference may be attributed to the sample dependent
disorder. On the other hand, our measurements were done under a weak
continous pump-probe optical excitation that can contribute to depinning
and helps to trigger the transition at higher $T$. 

Overall, the observed effect of Se substition on the H phase is two
fold. Firstly, the transition temperature T$_{H}$ increases slightly
or not at all, with increasing $x$ as a result of increasing tensile
strain\cite{LiuAPL} exerted by the larger Se substitution. We can
qualitatively compare the chemical strain effect of Se ion substitution
with in-plane differential strain experiments with thin crystals of
1T-TaS$_{2}$ on different substrates, where a tensile strain on cooling
was found to empirically increase $T_{H}$. The two trends thus agree
qualitatively.

The second effect is that the threshold fluence increases with increasing
$x$. Pinning by the Se dopants would be expected to have an opposite
effect in stabilizing the domain structure created in the switching
process. Another possibility is that the absorption coefficient is
changing with Se doping, changing the effective carrier density. The
photon absorption near 1.5 eV is related to a charge-transfer excitation
between metal and chalcogen, and is close to the absorption edge for
this transition. It is therefore concievable that the chalcogen substitution
can have an effect on the imaginary part of the dielectric constant
in this region. Unfortunately, presently ellipsometric data are not
available to confirm or disprove this hypothesis. The effect of disorder
may also have an influence if doping is associated with traps (e.g.
on interstitials), thus reducing the photinduced carrier density.
A threfold increase in threshold fluence would require a significant
number of traps, which we consider unlikely. 

The frequency of the AM in the C phase does not show a systemtic $x$-dependence
while the corresponding mode frequency in the H-phase shows a small,\footnote{An order of magnitude smaller than $\gamma$.}
but detectable, dependence {[}see Fig. \ref{fig:Temp-H-State} c){]}.
The absence of the shift in the C-phase indicates that the AM mode
eigenvector does not include significant (S,Se) site displacements.
Since the corresponding mode observed in the H-phase hardens slightly
with increasing $x$ the shift can not be related to the larger Se
mass. Considering that the broadening of the modes is the most likely
inhomogenous and $\gamma$ is an order of magnitude larger {[}see
Fig. \ref{fig:Temp-H-State} b){]} than the shift the origin of the
shift can be attributed to a doping dependence of the Se inhomogeneity.

Moreover, in the $x=0.5$ sample a notable softening of $\sim0.03$
THz in comparison to the $x\ne$0.5 samples is observed in both phases.
This suggest a possibility of a Se ordering at this doping since $x=0.5$
correponds to a commensurate 25\% filling of the triangular (S,Se)
site lattice. The ordering, however, can not be long range since the
mode linewidth is the largest at this doping. Nevertheles, the slight
increase of the the H-phase stability at $x=0.5$ could be tentatively
linked to enhanced pinning of the H-phase domain walls by partially
ordered rows of Se ions.

\section*{Conclusions}

We demonstrated a successful transition of 1\textit{T}-TaS$_{2-x}$Se$_{x}$
system to the hidden photo-induced state over a major portion of the
Mott-phase region ($x\leq0.6$). The H-state transition is triggered
by an ultrafast laser quench similarly to the pristine $1T$-TaS$_{2}$,\cite{StojchevskaScience}
but at fluences that significantly increase with increasing $x$ from
$\mathcal{F}_{\mathrm{H}}\sim1$ mJ/cm$^{2}$ at $x=0$ to $\mathcal{F}_{\mathrm{H}}\sim4$
mJ/cm$^{2}$ at $x=0.5$. The high temperature stability of the H-phase
is not significantly influenced by the Se doping despite a much larger
effect on the stability of the equilibrium C phase. An exception is
$x=0.5$ Se doping where the slight increase of the the H-phase stability
concurent with the AM mode softening could be tentatively linked to
partial Se-ion ordering. 
\begin{acknowledgments}
The authors acknowledge the financial support of Slovenian Research
Agency (research core funding No-P1-0040) and European Research Council
Advanced Grant TRAJECTORY (GA 320602) for financial support. L. S.
would also like to acknowledge supported by MIZŠ\&ESS funds, ULTRA-MEM-Device
project, 2014-2015.
\end{acknowledgments}

\bibliographystyle{apsrev4-1}
\bibliography{biblio}

\end{document}